# Complex macroscale structures formed by the shock processing of amino acids and nucleobases – Implications to the *Origins of life*

V S Surendra[1], V Jayaram[2], S Karthik[1], S Vijayan[1], V Chandrasekaran[3], R Thombre[4], T Vijay[5], B N Raja Sekhar[6], A Bhardwaj[1], G Jagadeesh[7], K P J Reddy[7], N J Mason[8], B Sivaraman[1],*

[1]Atomic Molecular and Optical Physics Division, Physical Research Laboratory, Ahmedabad, India.
[2]Solid State & Structural Chemistry Unit, Indian Institute of Science, Bangalore, India.
[3]Department of Chemistry, Vellore Institute of Technology, Vellore, India.
[4]Department of Biotechnology, Modern College of Arts and Science, Pune, India.
[5]Department of Chemistry, Indian Institute of Technology – Gandhinagar, Gandhinagar.
[6]Atomic and Molecular Physics Division, BARC at RRCAT, Indore, India.
[7]Department of Aerospace Engineering, Indian Institute of Science, Bangalore, India.
[8]School of Physical Sciences, University of Kent, United Kingdom.

## Abstract

The building blocks of life, amino acids and nucleobases, are believed to have been synthesized in the extreme conditions that prevail in space starting from simple molecules containing hydrogen, carbon, oxygen and nitrogen. However, the fate and role of amino acids and nucleobases when they are subjected to similar processes largely remains unexplored. Here we report, for the first time, that shock processed amino acids and nucleobases tend to form complex macroscale structures. Such structures are formed on timescales of about 2 ms. This discovery suggests that the building blocks of life could have polymerized not just on Earth but on other planetary bodies. Our study also provides further experimental evidence for the 'threads' observed in meteorites being due to assemblages of (bio)molecules arising from impact induced shocks.

* Corresponding author - bhala@prl.res.in

The study of 'impacts' is crucial in our understanding of planet formation, evolution and the physico-chemical processes that lead to molecular synthesis in the interstellar medium and in proto-planetary systems. Bolide impacts (asteroid and meteorites) are also expected to have played a key role in the evolution of life on Earth either through mass extinction events (such as one leading to end of dinosaur age) or through routes by which prebiotic material was transported to Early Earth. The feasibility of panspermia (transport of life between planetary bodies) is also dependent upon impact studies. The observation of large scale craters on the surface of the icy satellites reminds us of the role of impact processes in planetary and lunar evolutions whilst many cometary bodies appear to be the result of collision of constituent bodies. Such impacts release significant amounts of energy and therefore may provide pathways for large scale molecular synthesis.

The collection of meteorites (including those identified as originating from Mars) has led to renewed interest in the physical and chemical changes induced during impact and shock of extra-terrestrial material into the Earth's atmosphere and impact on terrestrial surface. Experiments have shown that the impact and shock processes on simple molecules leads to the synthesis of building blocks of life, such as amino acids ((*1, 2*) (*3, 4*)). Apart from impacts, irradiation processes on such simple molecule frozen on dust grains were also found to synthesize amino acids, and some in cases, the nucleobases (uracil, cytosine Thymine (*5-7*)) and sugars (ribose(*8*)) may be formed. So eventually, all the building blocks of life can be derived starting from the simplest of molecules. However, impact associated shock processing of amino acids, nucleobases (along with sugars) in extreme conditions largely remains unexplored (*9, 10*). Blanke et al(*11*), used an impactor to process an amino acid 'soup' and observed peptide bonds to be present in the shocked solution. However, since the Urey-Miller experiment and until the recent impact induced formation of complex molecules (*3*) the role of such molecules in extreme conditions and in the origin of life, is not known.

We have subjected amino acids and nucleobases to conditions that prevail in the bolide impact events using a shock tube (Fig S1) that can reach temperatures up to 8000 K for about 2 ms (Fig S2). The shocked samples collected from the shock tube were then subjected to Scanning Electron Microscope (SEM) and Transmission Electron Microscope (TEM) imaging. We have observed compelling evidence on the formation of polymers containing complex macrostructures. These experiments therefore provide evidence for new route of formation of complex macroscale structures from the building blocks of the *Origins of life*.

The experiments were carried out using an 8-meter-long shock tube that can be pumped down to base pressure of 10-4 mbar on the driven side (shocked region) and very high pressures, up to 60 bar are achieved on the driver side (shock generation region). In all the experiments Argon (Ar) was used in the driven section and Helium (He) is used in the driver section. An aluminium diaphragm (~ 5 mm) is used to separate the high and low pressure regions. The reaction chamber is placed at the end of the driven section of the shock tube separated from it by a valve, where the samples are loaded for processing. Samples containing (up to 20) amino acids, in equal proportions (w/w) mixture, and nucleobases (in equal proportions w/w mixtures of AGCT / AGCU), are pre-loaded into the reaction chamber, before pumping down the driven section, kept at room temperature before shock processing. Upon diaphragm rupture due to the very high pressure on the driver side the shock front developed processes the solid samples in the reaction chamber. The diaphragm thickness and the pressure in the driven side determines the final shock temperature. In the present experiment the shock temperature rises to few thousands of kelvin (~ 8000 K) for 2 ms. The valve separating the reaction chamber and shock tube was immediately closed after the shock wave had passed so that sufficient samples were obtained after every experiment. The samples collected after shock processing were imaged using both SEM and TEM.

**Results**

The main focus of this work was to search for the formation of polymers. Imaging of the residues obtained from shocked samples collected from the reaction chamber section of the shock tube were carried out. Since previous studies had not reported any imaging analysis of the residues obtained. SEM imaging of the unshocked sample were found to show solid clumps of raw sample, sizes ranging up to 10 μm. However, the shocked amino acid, glycine, revealed an entirely different textured (Fig.1), a globule of ~ 45 to 50 μm diameter, with smooth texture and spongy appearance. Interestingly, a ribbon/thread structure, stretching to a few 10's of a micron, was also observed (Fig.1). While by adding another amino acid, glutamic acid, in equal proportions with glycine showed different macrostructures (Fig. 2), flower petals and shorter bunched threads having been formed. Similarly, the shock processed mixture of asparagine and glutamic acid samples presented different structures (Fig. 2). Further experiments made by adding two more amino acids, a mixture of four (Table.1), in equal proportions resulted in a dark black residue that was very sticky. SEM imaging of this last sample showed threads (Fig. 3) and, a much more surprising result, the formation of a cylindrical structure (Fig. 3), a few microns in diameter.

Increasing the number of amino acids in the mixture to eight (Table 1), mixed in equal proportions w/w ratio, resulted in a shocked sample that was again observed to be very sticky with the structures in SEM imaging observed to be solid clumps along with twisted and a few cylindrical threads. By further increasing the number of amino acids to between eighteen and twenty in the mixture, a thick black sticky residue resulted and the structures made were threads and ribbons; twisted and cylindrical (Fig 4a). Upon closer inspection, we could clearly see that the threads were made of small (about a micron size) features. By reducing the shock temperature, the number of threads formed was found to increase. The length of the threads formed was quite surprising as they spanned more than 1 mm. In addition, portions of the thread observed were found to be buried in the solid clumps. Even spherical particles were observed to have formed in all the shock-processed samples, a feature that is characteristic of shock processed micron size samples. The twisted threads were observed to split (Fig 4a), an indication of an even more complex structure. Most of the threads were observed to be solid; nevertheless, tubular structures were also found (Fig 4b). So a variety of structures were observed when many amino acids are mixed together and are subjected to extreme conditions. TEM analysis of the shocked (18 mixture) amino acids revealed, membrane like structure with branching features being clearly observed with fine threads running throughout (Fig S3).

When nucleobases were subjected to similar shock conditions, a blackish residue was obtained. The SEM images of the residue showed compelling evidence for the formation a highly twisted thread structure along with solid clumps (Fig 5). Similar structures were observed while ribose was added to nucleobases.

**Conclusion**

This study provides the clear evidence for the formation of complex macrostructures when the molecular building blocks of life are subjected to impact conditions that commonly prevail in the solar system. The different structures that are made, within a few milliseconds, were found to depend on the different mixtures of amino acid. Similarly, complex structures observed after shocking nucleobases also suggest such conditions with the right prebiotic mixture (consisting of amino acids, nucleobases, fatty acids, sugars and phosphates), could have favoured the formation of further more complex assemblies. The threads formed, especially the observed tubular structures compelling evidence that impact events may have played an important role in making complex (bio)structures that are imperative for the *origin of life*. The thread like features reported in the meteorites, which were initially suggested to be fossilized life form (*12, 13*), could actually be the assemblages of biomolecules, such as amino acids etc., that have undergone shock processing during impact events.

**Acknowledgements**

All the authors thank the support from Physical Research Laboratory (Department of Space, Government of India), Indian Institute of Science (India), Sir John Mason Academic Trust (UK) and Indian Institute of Technology – Gandhinagar (India) during the course of the experiments. Authors specially acknowledge the nanoscience imaging facilities at IISc - Bangalore, ILS – Ahmedabad University and SICART, Gujarat, for the SEM and TEM imaging. BS personally thank Dr Akshay and Prof Dipshikha, Department of Microbiology and Cell Biology, Indian Institute of Science, Bangalore, for providing the amino acid samples for the first experiment.

**Competing financial interest**

The authors do not have any competing financial interest.

**Author contributions**

VSS, VJ, SK, SV, VC, RT, BS performed the experiments. VSS, VJ, SK, TV, SV, BNRS, BS performed SEM and TEM imaging. All the authors took part in discussions, contributed to the analysis and manuscript preparation. BS conceived the idea and supervised the study.

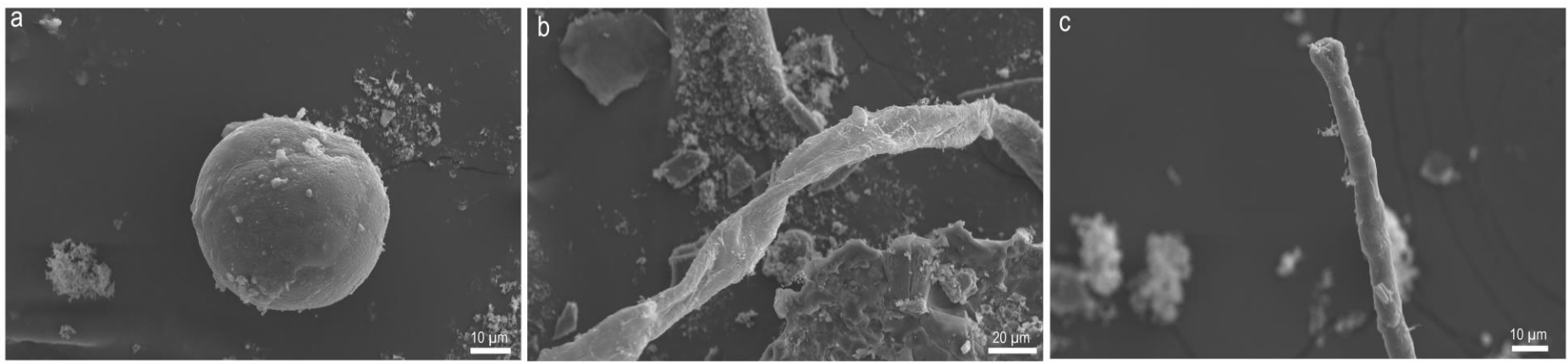

Fig 1: SEM images of the residues obtained after shock processing glycine. (a) Glycine globule, (b) fine thread feature, and (c) cylindrical thread feature.

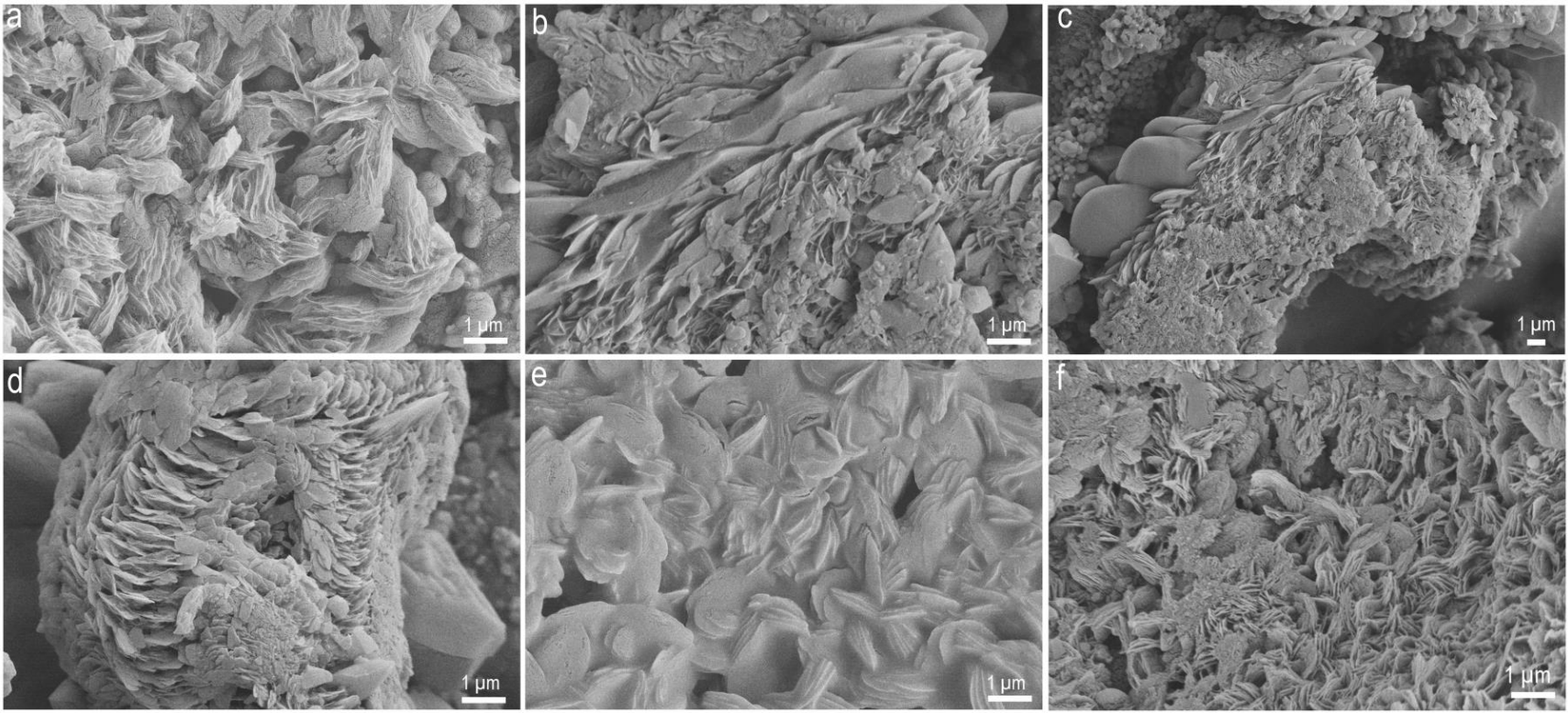

Fig 2: SEM images of residues obtained after shock processing of two amino acid mixtures (Table 1). (a,b & c) short feather like features, (d) shows the formation of ordered scale features, (e) rose petal like features and (f) short thread like features.

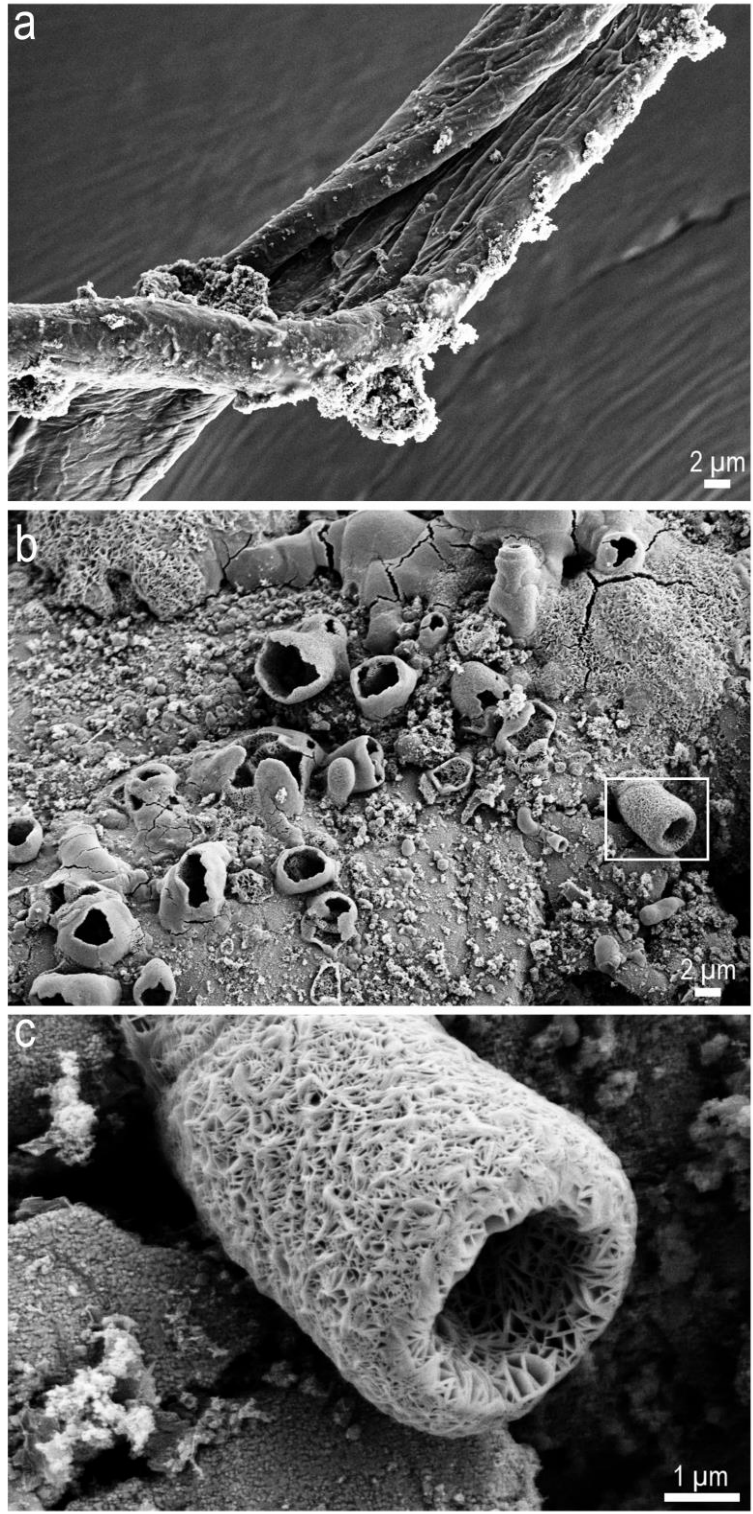

Fig 3: SEM images of residues obtained after shock processing of equal proportions of four amino acids (Table 1). (a) thick thread feature containing fine threads running all along and (b) shows many porous features and (c) the porous cylindrical feature.

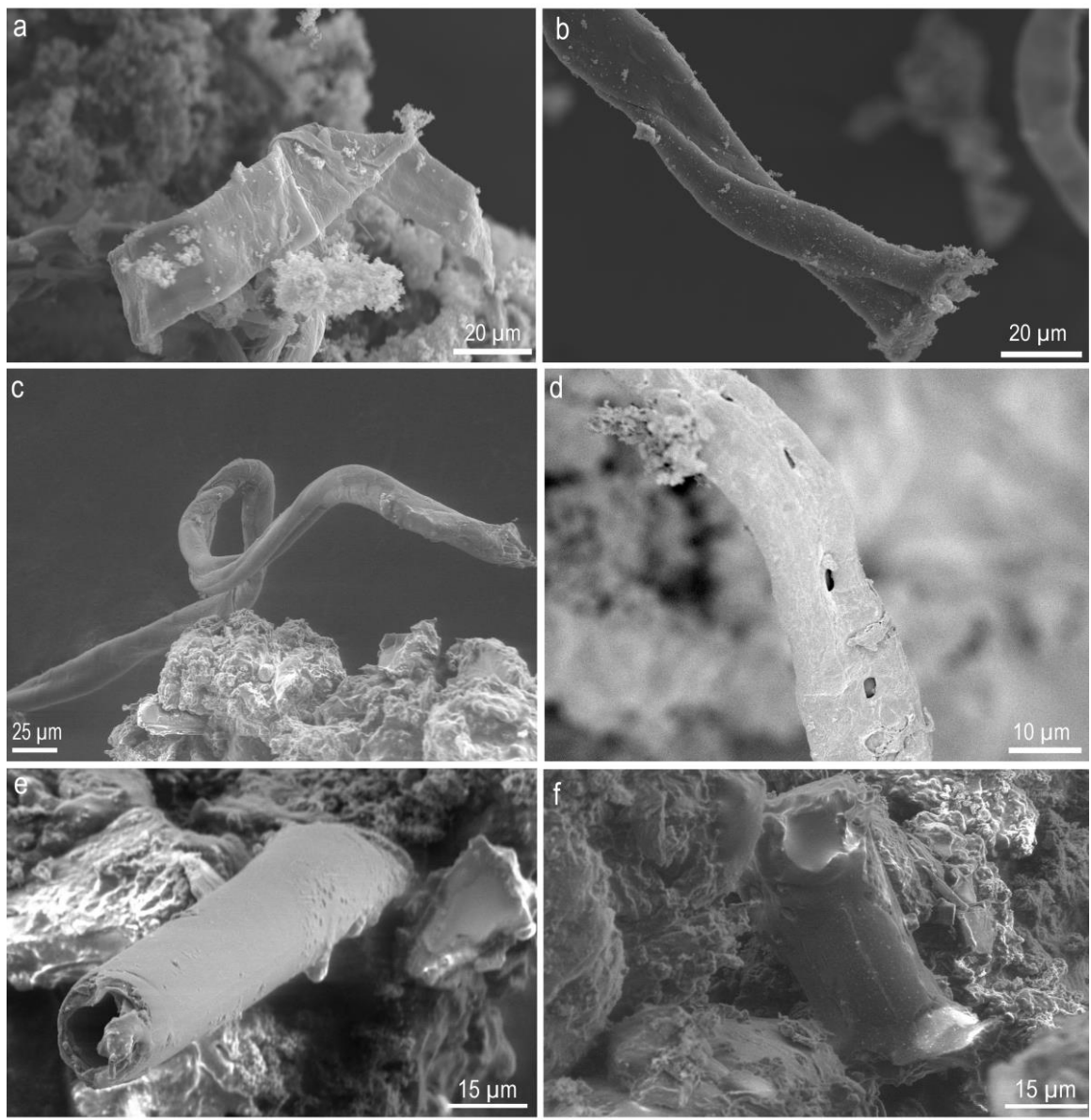

Fig 4a: SEM images of residue obtained after shocking processing of 18 amino acids (Table 1). Several features including (a) thin ribbon, (b) twisted and (c) branching threads and (d, e & f) tubes are observed.

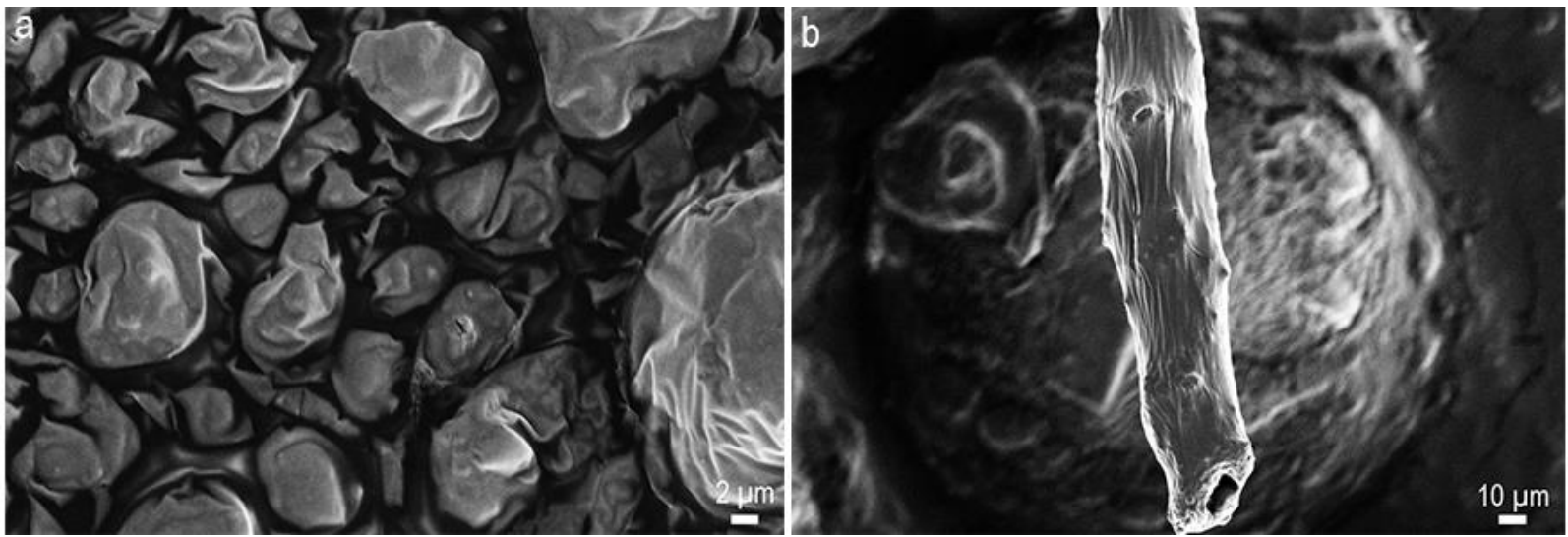

Fig 4b: SEM images of residue obtained after shock processing of 20 amino acid mixtures (Table 1). (a) sheet like features and (b) tubular structure are also observed.

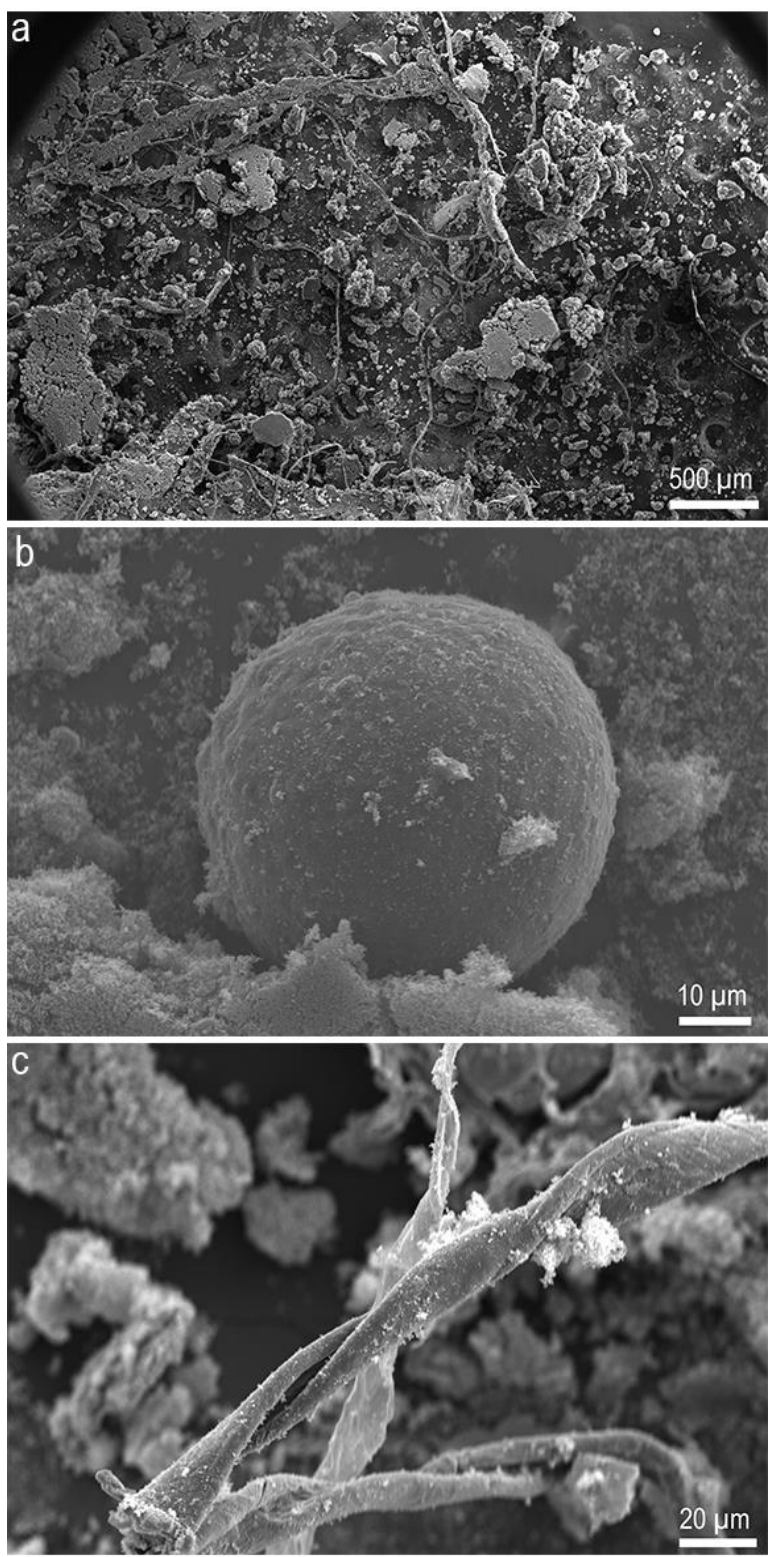

Fig 5: SEM images of shock processed mixture of nucleobases (Table 1) showing (a) bulk sample containing many threads, (b) globule feature and (c) twisted thread.

**Table 1: Shock parameters for different experiments**

| S. No. | Run No. | Quantity (g) | Test gas pressure (Bar) | Shock Mach number | Reflected shock temperature (K) | Reflected shock pressure (Bar) | Features seen in SEM |
|---|---|---|---|---|---|---|---|
| | | | | **Glycine** | | | |
| 1 | 675 | 0.18 | 0.05 | 5.95 | 8088 | 28.4 | Folded threads |
| 2 | 677 | 0.18 | 0.154 | 4.67 | 5024 | 25.1 | NA |
| 3 | 825 | 0.4 | 0.075 | 5.18 | 6157 | 25.3 | Folded thread |
| 4 | 826 | 0.4 | 0.05 | 5.55 | 7047 | 21.8 | Globules |
| 5 | 827 | 0.4 | 0.194 | 4.08 | 3871 | 16.3 | Flower structure |
| | | | | **Two mixture - amino acids (Glycine & Glutamic acid)** | | | |
| 6 | 823 | 0.4 | 0.075 | 5.25 | 6321 | 19.6 | Petals structure |
| 7 | 828 | 0.4 | 0.16 | 4.67 | 5024 | 16.8 | Petals and buds structure |
| 8 | 829 | 0.4 | 0.029 | 5.80 | 7679 | 18.8 | Threads and petals |
| | | | | **Two mixture - amino acids (Asparagine & Glutamic acid)** | | | |
| 9 | 696 | 0.2 | 0.25 | 4.21 | 4118 | 25.4 | NA |
| 10 | 711 | 0.8 | 0.25 | 4.36 | 4413 | 25.1 | Threads |
| 11 | 726 | 0.5 | 0.25 | 4.27 | 4228 | 24.0 | NA |
| 12 | 731 | 0.5 | 0.25 | 2.53 | 1574 | 3.2 | Cubes |
| 13 | 830 | 0.4 | 0.075 | 5.20 | 6197 | 22.8 | Thread, sheet of woven petals |
| 14 | 831 | 0.4 | 0.16 | 4.67 | 5024 | 22.0 | Small folded threads |
| 15 | 832 | 0.4 | 0.029 | 5.80 | 7679 | 22.2 | Petals |
| colspan="8" | Four mixture - amino acids (Lysine, Aspartic acid, Arginine, Glutamic acid) |
| 16 | 694 | 0.2 | 0.1 | 5.34 | 6633 | 32.9 | Thread, tube like |
| 17 | 695 | 0.2 | 0.25 | 3.92 | 3594 | 19.5 | Folded thick flat threads, woven tube |
| 18 | 712 | 0.5 | 0.25 | 4.02 | 3775 | 19.5 | Small folded flat thread, root like feature |
| 19 | 819 | 0.5 | 0.20 | 4.32 | 4329 | 22.3 | Threads |
| 20 | 820 | 0.5 | 0.20 | 4.18 | 4020 | 17.6 | Cracked sheets and balls |
| 21 | 833 | 0.5 | 0.075 | 5.09 | 5961 | 24.4 | Folded threads |
| 22 | 834 | 0.5 | 0.025 | 5.86 | 7850 | 20.1 | Threads |
| colspan="8" | 18 mixture - amino acids (Glycine, Alanine, Valine, Leucine, Proline, Serine, Aspartic acid, Glutamic acid, Cysteine, Methionine, Asparagine, Glutamine, Arginine, Histidine, Phenylalanine, Lysine, Tyrosine, Tryptophan) |
| 30 | 676 | 0.186 | 0.055 | 5.63 | 7284 | 28.2 | Folded threads |
| 31 | 678 | 0.12 | 0.25 | NA | NA | NA | Full threads |
| 32 | 692 | 0.18 | 0.25 | 7.07 | 3851 | 22.2 | NA |
| 33 | 702 | 1.08 | 0.25 | 3.97 | 3683 | 21.3 | NA |
| 34 | 704 | 0.9 | 0.25 | 3.93 | 3611 | 16.8 | Long thick thread |
| 35 | 705 | 1.016 | 0.5 | 3.35 | 2666 | 16.3 | NA |
| 36 | 814 | 0.5 | 0.25 | 4.22 | 4139 | 24.7 | Threads, Tube |
| 37 | 815 | 0.4 | 0.2 | 4.13 | 3971 | 15.5 | Small thread |
| 38 | 835 | 0.54 | 0.075 | 5.29 | 6405 | 22.2 | Long folded thick threads |
| 30 | 836 | 0.54 | 0.1 | 4.86 | 5429 | 22.8 | Small threads |
| | | | | | | | |
| **20 mixture - amino acids** <br><br> **(Glycine, Alanine, Valine, Leucine, Isoleucine, Threonine, Proline, Serine, Aspartic acid, Glutamic acid, Cysteine, Methionine, Asparagine, Glutamine, Arginine, Histidine, Phenylalanine, Lysine, Tyrosine, Tryptophan)** ||||||||
| 40 | 727 | 0.5 | 0.25 | 3.77 | 3330 | 12.7 | Flat thread, tube |
| 41 | 728 | 0.5 | 0.25 | 4.05 | 3813 | 19.7 | NA |
| **AGCT (nucleobases)** ||||||||
| 42 | 690 | 0.16 | 0.102 | 5.27 | 6363 | 28.8 | Threads |
| **Mixture – AGCT (nucleobases) and Ribose** ||||||||
| 43 | 691 | 0.12 | 0.075 | 5.49 | 6901 | 34.2 | Threads |
| **Mixture - AGCT, Ribose, Potassium phosphate** ||||||||
| 44 | 839 | 0.5 | 0.075 | 5.25 | 6321 | 22.4 | Folded ribbons |

NA- Sample used for other analysis

**Supplementary material**

Methodology

The shock tube used in the current research is the material shock tube (MST1, *(10)*) in the Department of Solid State Structural Chemistry Unit, Indian Institute of Science, Bangalore, India. It is an 80 mm diameter shock tube composed of 2 m long driver section and 6 m long driven section separated by a metallic diaphragm. Generally, aluminium diaphragms up to about 5 mm thickness are used with appropriate grooves to guide the proper pressure bursting to produce shock waves of required strength. After cleaning the shock tube using acetone, the diaphragm of desired thickness is placed between driver section and driven section. To obtain different reflected shock temperatures, diaphragms of different thickness are used. The sample holder for the study of shock wave interactions with test samples is attached to the end of the shock tube through a manually operated gate valve. The test sample is spread over a thin plate parallel to the flow in the middle of the shock tube and the interaction with the shocked gas behind the reflected shock wave occurs in the 30 cm long reaction chamber downstream. A schematic diagram of MST1 is shown in Fig.S1

The amino acid mixture is spread on sample holder. The driven section is purged two times with ultra-high pure Argon (99.999%) then pumped to high vacuum up to $2 \times 10^{-4}$ mbar to avoid gas impurities. After sample loading pumping is performed at slower rate to avoid pumping and spreading of the sample. The driven section is filled with ultra-high pure Argon (99.999%) up to pressure of 0.1 bar. The driver section is rapidly filled with high pressure Helium gas at very high mass flow rate until the diaphragm ruptures. The sudden rupture of the diaphragm generates a shock wave which travels through the driven section and is reflected back from the end flange of driven section. The ball valve next to reaction chamber is closed immediately after the rupture of the diaphragm. The high pressure inside the reaction chamber is brought to equilibrium by slowly exhausting in to atmosphere. Then the solid residue left in the reaction chamber is collected for further analysis.

Three pressure sensors (PCB Piezotronics 113B22) surface mounted at three different locations on the driven section are used to obtain a pressure signal with help of Tektronix digital oscilloscope (TDS2014B). The shock speed and Mach number is calculated by finding the time taken (Δt) by the shock wave to cross the location (Δx) of two pressure sensors with the help of pressure signal recorded. Typical pressure signals recorded by oscilloscope are shown in the Fig S2. The temperature and the pressure behind the reflected shock wave are calculated using following one dimensional normal shock equations known as R-H relations, for the shock Mach number obtained experimentally as follows,

$$V_S = \frac{\Delta x}{\Delta t} \; ; \; M_S = \frac{V_S}{a} = \frac{V_S}{\sqrt{\gamma R T_1}} \qquad (1)$$

$$\frac{T_5}{T_1} = \frac{\{2(\gamma-1)M_S^2+(3-\gamma)\}\{(3\gamma-1)M_S^2-2(\gamma-1)\}}{(\gamma+1)^2 M_S^2} \qquad (2)$$

$$\frac{P_5}{P_1} = \left[\frac{2\gamma M_s^2-(\gamma-1)}{(\gamma+1)}\right]\left[\frac{-2(\gamma-1)+M_s^2(3\gamma-1)}{2+M_s^2(\gamma-1)}\right] \quad (3)$$

where $\gamma$ is specific heat ratio of test gas argon, R is gas constant, a is speed of sound in test gas argon and $T_1$ is the temperature of test gas. The reflected gas pressure $P_5$ and temperature $T_5$ are related to the measured shock Mach number Ms. Mach numbers and the shock temperature ($T_5$) and pressure ($P_5$) for each experiment are listed in Table 1.

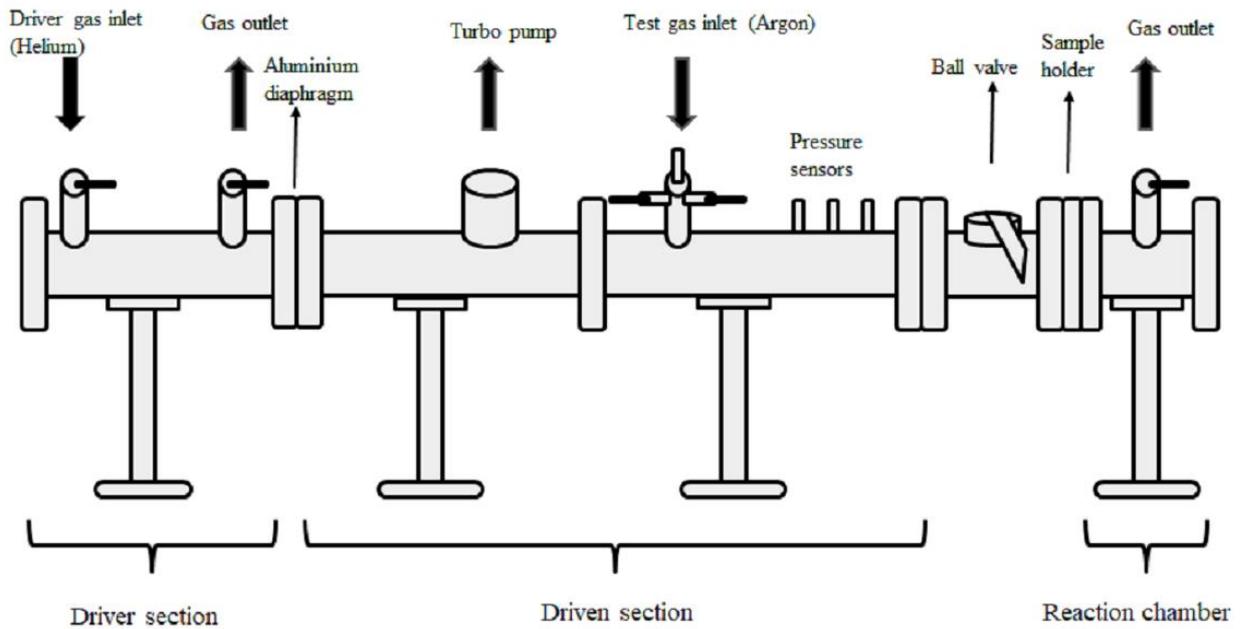

Fig S1: Schematic diagram of material shock tube (MST1).

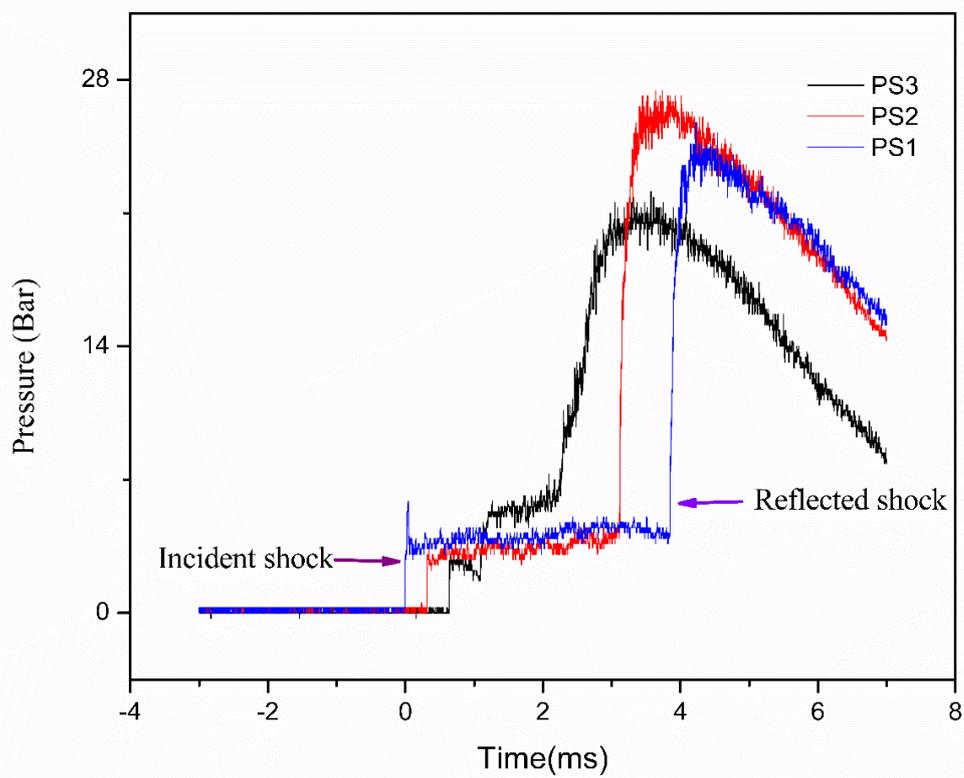

Fig S2: Typical pressure signal recorded using oscilloscope.

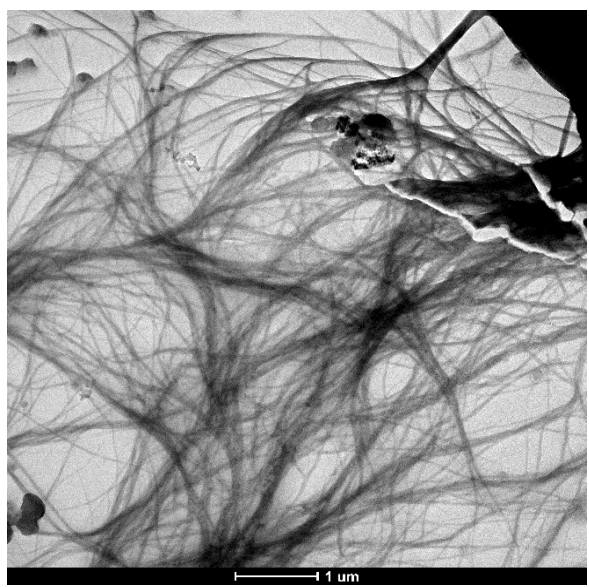 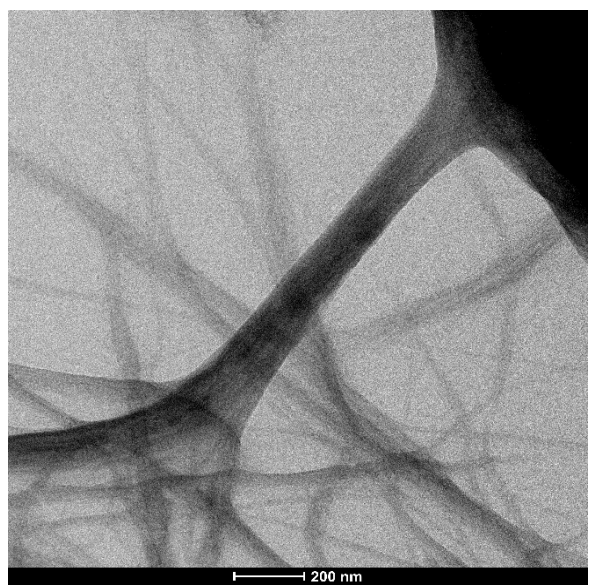

Fig S3: TEM image of shocked 18 amino acid mixture (Table 1).